\documentclass[twocolumn,prd,showpacs,a4paper]{revtex4}
\begin{document}
\title{Identity of the imaginary-time and real-time thermal propagators for scalar bound states in a one-generation Nambu-Jona-Lasinio model 
\\
}
\author{Bang-Rong Zhou}\email[Electronic mailing address: ]{zhoubr@163bj.com}

\affiliation{  
Department of Physics,the Graduate School of Chinese Academy of Sciences,Beijing 100039, China \\
}
\begin{abstract}
By rigorous reanalysis of the results, we have proven that the propagators at finite 
temperature for scalar bound states in one-generation fermion condensate scheme of 
electroweak symmetry breaking are in fact identical in the imaginary-time and the 
real-time formalism. This dismisses the doubt about possible discrepancy between the 
two formalisms in this problem. Identity of the derived thermal transformation 
matrices of the real-time matrix propagators for scalar bound states without and with 
chemical potential and the ones for corresponding elementary scalar particles shows 
similarity of thermodynamic property between the two types of particles. Only one 
former inference is modified, i.e. when the two flavors of fermions have unequal 
nonzero masses, the amplitude of the composite Higgs particle will decay instead grow 
in time.  
\end{abstract}
\maketitle
Finite temperature field theory has been extensively researched owing to its 
application to evolution of early universe and phase transition of the nuclear matter 
[1-3]. However, the demonstration of the equivalence of its two formalisms - the 
imaginary-time and the real-time formalism [2] -is often a puzzling problem and has 
been extensively concerned [4-7]. In a recent research on the Nambu-Goldstone 
mechanism [8] of dynamical electroweak symmetry breaking at finite temperature based 
one-generation fermion condensate scheme ( a Nambu-Jona-Lasinio (NJL) model [9]), 
we calculate the propagators for scalar bound states from four-point amputated 
functions in the two formalisms which seem to show different imaginary parts in their 
denominators [10]. This difference is quite strange considering that the analytic 
continuation used in Ref. [10] of the Matsubara frequency in the imaginary-time 
formalism was made as the way leading to the causal Green functions obtained 
in the real-time formalism and that in the fermion bubble diagram approximation, the 
calculations of the four-point amputated functions in a NJL model may be effectively 
reduced to the ones of two-point functions ( though they have been subtracted through 
use of the gap equation [11,12]), and it is accepted that a two-point function should 
be equivalent in the two formalisms. Therefore, we have to reexamine the whole 
calculations in Ref. [10]. We eventually find that the origin of the above difference 
is that we did not rigorously keep the general form of the analytic continuation and 
not explicitly separate the imaginary part of the zero-temperature loop integral from 
relevant expressions. In this paper we will use the main results of the propagators 
for scalar bound states obtained in Ref. [10], but correct some expressions which 
were not exact enough and complete rigorous derivation to the final results in both 
the formalisms. Unless specified otherwise, all the denotations will be the same as 
those in Ref. [10]. 

First we discuss the neutral scalar bound state  $\phi^0_S$.
In the imaginary-time formalism, by means of the analytic continuation of the 
Matrubara frequency $\Omega_m $ of $\phi^0_S$  
\begin{equation}
-i\Omega_m \longrightarrow p^0 + i\varepsilon \eta(p^0), \ \ \ \varepsilon =0_+,\ \ \ \eta(p^0)=p^0/|p^0|
\end{equation}
we obtain the propagator for $\phi^0_S$ [10]
\begin{widetext}
\begin{equation}
\Gamma_I^{\phi_S^0}(p)=-i\sum_{Q}m_Q^2/\sum_{Q}(p^2-
4m_Q^2+i\varepsilon)m_Q^2[K_Q(p)+H_Q(p)-iS_Q^I(p)],  
\end{equation}
where 
\begin{equation}
S_Q^I(p)=\eta(p^0)4\pi^2d_Q(R)\int
\frac{d^4l}{{(2\pi)}^4}\delta(l^2-m_Q^2)\delta[(l+p)^2-m_Q^2] \left[\sin^2\theta(l^0,\mu_Q)\eta(l^0+p^0)+\sin^2\theta(l^0+p^0, \mu_Q)\eta(-l^0)\right].
\end{equation}
\end{widetext}
Eq. (3) is somehow different from Eq. (3.29) in Ref.[10], but is more general 
since in its derivation the original form (1) of the analytic continuation is always 
kept; instead, in Ref. [10], $\eta(p^0)$ was replaced by 
$\eta(\omega_l-\omega_{l+p})$ or $\eta(\omega_{l+p}-\omega_l)$ or $\pm 1$ depending 
on the sign of the pole of $p^0$ in a term. As will be seen later, Eq.(3) is more 
suitable to the proof of equivalence of two formalisms. It is indicated that , owing 
to the factors $\eta(l^0+p^0)$ and $\eta(-l^0)$ in the integrand, $ S_Q^I(p)$ in Eq. 
(3) does not contain any singularity when $p\to 0$. The zero-temperature loop integral
$K_Q(p)$ is complex and can be written by
\begin{equation}
K_Q(p)=K_{Qr}(p)+K_{Qi}(p).
\end{equation}
By applying the residue theorem of complex $l^0$ integral to the first formula in 
Eq. (3.27) in Ref.[10] we can obtain the imaginary part of $K_Q(p)$
\begin{eqnarray}
K_{Qi}(p)&=&\frac{d_Q(R)}{16\pi^2}\int\frac{d^3l}{\omega_{Ql}\omega_{Ql+p}}
   \left[\delta(p^0+\omega_{Ql}+\omega_{Ql+p})\right.\nonumber \\
&&\left.+\delta(p^0-\omega_{Ql}-\omega_{Ql+p})\right].
\end{eqnarray}
The $\delta$-functions in Eq.(5) ensure that $K_{Qi}(p)\neq 0$ only if $p^2\geq 
4m_Q^2$. From Eq. (3) we can derive
\begin{equation}
S^I_Q(p)=R_Q(p)\sinh(\beta|p^0|/2)+K_{Qi}(p),
\end{equation}
where $R_Q(p)$ is given by Eq.(6.5) in Ref. [10]. Hence the physical causal propagator
 (2) for $\phi_S^0$ in the imaginary-time formalism becomes
\begin{eqnarray}
\Gamma_I^{\phi_S^0}(p)&=&-i\sum_{Q}m_Q^2/\left\{[k_r+h-ir\sinh(\beta|p^0|/2)]p^2\right.\nonumber\\
&&\left.-4[\tilde{k}_r+\tilde{h}-i\tilde{r}\sinh(\beta|p^0|/2)]\right\},
\end{eqnarray}
where $k_r$, $h$, $\tilde{k}_r$, $\tilde{h}$ are defined by Eq. (3.32) and $r$, 
$\tilde{r}$ by Eq. (6.3) in Ref.[10]. 
On the other hand, in the real-time formalism, we must explicitly separate the 
imaginary part of $K_Q(p)$ as Eq.(4) and this operation was ignored in Ref. [10],
thus the correct matrix propagator $\Gamma^{\phi_S^0ba}(p) \ (b,a=1,2)$ can be 
obtained from Eq. (6.2) in Ref.[10] by the replacements
\begin{eqnarray}
k&\rightarrow &k_r, \ \ \tilde{k} \rightarrow \tilde{k}_r,\nonumber \\
s&\rightarrow &s'=\sum_{Q} m_Q^2[S_Q(p)-K_{Qi}(p)]=r\cosh(\beta p^0/2),\nonumber \\
\tilde{s}&\rightarrow &\tilde{s}'=\sum_{Q} 
m_Q^4[S_Q(p)-K_{Qi}(p)]=\tilde{r}\cosh(\beta p^0/2),
\end{eqnarray}
where the relation
\begin{equation}
S_Q(p)-K_{Qi}(p)=R_Q(p)\cosh(\beta p^0/2)
\end{equation}
has been used. Correspondingly, we will have the replacement
\begin{eqnarray}
S/R&=&(p^2s-4\tilde{s})/(p^2r-4\tilde{r})\rightarrow \nonumber \\ 
S'/R&=&(p^2s'-4\tilde{s}')/(p^2r-4\tilde{r})=\cosh(\beta p^0/2).
\end{eqnarray}
Applying Eqs. (8) and (10) to Eq. (6.10) in Ref.[10], we find that the physical 
propagator $\Gamma_R^{\phi_S^0}(p)$ for $\phi_S^0$ in the real-time formalism is 
identical to the one in the imaginary-time formalism expressed by Eq.(7), i.e. 
$\Gamma_R^{\phi_S^0}(p)$=$\Gamma_I^{\phi_S^0}(p)$. In addition, the thermal 
transformation matrix $M_S$ which diagonalizes the matrix propagator  
$\Gamma^{\phi_S^0ba}(p) \ (b,a=1,2)$ will be reduced to
\begin{eqnarray}
M_S&=&\left(\matrix{\cosh \theta_S & \sinh \theta_S\cr
                  \sinh \theta_S &  \cosh \theta_S\cr}\right),
\nonumber \\
\sinh\theta_S&=&\left[\frac{1}{\exp(\beta |p^0|)-1}\right]^{1/2}.
\end{eqnarray}
Hence $M_S$ is identical to the thermal transformation matrix of the matrix propagator 
for an elementary neutral scalar particle [2]. This fact implies that that the scalar 
bound state $\phi_S^0$ and an elementary neutral scalar particle have the same 
thermodynamic property. \\
\indent By means of Eq.(4) which is different from $ K_Q(p)=K_{Qr}(p)-K_{Qi}(p)$ in 
Ref.[10] and Eq.(6), the equation to determine the squared mass of $\phi_S^0$ (Eq. 
(3.36) in Ref.[10]) will be changed into 
\begin{equation}
m^2_{\phi_S^0}=p^2_r=4\left.\frac{(\tilde{k}_r+\tilde{h})(k_r+h)+
               \tilde{r}r\sinh^2(\beta|p^0|/2)}
             {(k_r+h)^2+r^2\sinh^2(\beta|p^0|/2)}\right|_{p=p_r}.
\end{equation}
To reproduce the mass inequalities of $\phi_S^0$, we must determine the sign of 
$R_Q(p)$ in $r$ and $\tilde{r}$. In fact, $R_Q(p)$ given by Eq.(6.5) in Ref.[10] can 
be rewritten by
\begin{widetext}
\begin{eqnarray}
R_Q(p)
&=&2\pi^2d_Q(R)\int
\frac{d^4l}{{(2\pi)}^4}\delta(l^2-m_Q^2)\delta[(l+p)^2-m_Q^2]
\sin 2\theta(l^0,\mu_Q)\sin 2\theta(l^0+p^0, \mu_Q) \nonumber \\
&=&\frac{d_Q(R)}{32\pi^2}\int\frac{d^3l}{\omega_{Ql}\omega_{Ql+p}}\left\{
\frac{\delta(p^0-\omega_{Ql}+\omega_{Ql+p})-\delta(p^0-\omega_{Ql}-\omega_{Ql+p})}{\cosh[\beta(\omega_{Ql}+\mu_Q)/2]\cosh[\beta(p^0-\omega_{Ql}-\mu_Q)/2]} 
-(\omega_{Ql}\to -\omega_{Ql})\right\}.
\end{eqnarray}
\end{widetext}
The $\delta$-functions in first equality of Eq.(13) imply that 
\begin{equation}
R_Q(p)=0, \; \; {\rm when}\; 0\leq   p^2<4m_Q^2.
\end{equation}
Then from the second equality in Eq.(13), by means of the non-zero conditions of the 
four $\delta$-functions in the integrand that 
$\delta(p^0-\omega_{Ql}\pm\omega_{Ql+p}) \ 
(\delta(p^0+\omega_{Ql}\mp\omega_{Ql+p}))\neq 0$, if $p^2\geq 4m^2$ and $p^0>0 \ 
(p^0<0)$; $\delta(p^0-\omega_{Ql}+\omega_{Ql+p}) \ 
(\delta(p^0+\omega_{Ql}-\omega_{Ql+p}))\neq 0$, if $p^2<0$ and $p^0<0 \ (p^0>0)$,
we can see that when $\omega_{Ql}>\omega_{Ql+p}$ and $p^0>0$ $(p^0<0)$, only the 
first and the second (the third and the fourth) term are non-zero ones if $p^2\geq 
4m_Q^2$, but they cancel each other after integrating over the variable 
$\cos\chi=\stackrel{\rightharpoonup}{l}\cdot \stackrel{\rightharpoonup}{p}/
|\stackrel{\rightharpoonup}{l}||\stackrel{\rightharpoonup}{p}|$, thus there is no 
non-zero contribution in these cases; when $\omega_{Ql}<\omega_{Ql+p}$ and $p^0>0$ 
$(p^0<0)$, the non-zero terms will be the second (the third) one if $p^2\geq 4m_Q^2$
and the fourth (the first) one if $p^2<0$. Considering the signs of these terms and 
that $\stackrel{\rightharpoonup}{l}$ are integral variables we may reach the 
conclusion that $R_Q(p)<0$, if $p^2\geq 4m_Q^2$ and $R_Q(p)>0$, if $p^2<0$. In view 
of Eq.(14) we further have 
\begin{equation}
R_Q(p)<0 \ {\rm or} \ =0,\ \ {\rm if} \ \ p^2\geq 0.
\end{equation}
By this result together with $K_{Qr}(p)>0$ and $H_Q(p)>0$ [11], we will obtain from 
Eq.(12) the well-known mass inequalities
\begin{equation}
4(m_Q)^2_{\rm min}\leq m^2_{\phi_S^0} \leq 4(m_Q)^2_{\rm max}.
\end{equation}
The determination of the sign of $R_Q(p)$ will also change the sign of the imaginary
 part $p_i^0$ of the energy of $\phi_S^0$ when $0\neq m_U\neq m_D\neq 0$ obtained 
in Ref.[10]. In fact in present case $p_i^0\simeq b(p_r)/2p_r^0$ with
\begin{equation}
b(p_r)=4\left.\frac{[(\tilde{k}_r+\tilde{h})r-(k_r+h)\tilde{r}]\sinh\frac{\beta|p^0|}{2}}        {(k_r+h)^2+r^2\sinh^2\frac{\beta|p^0|}{2}}\right|_{p^2=p^2_r=m^2_{\phi_S^0}}.
\end{equation}
If we set $M_D=\alpha m_U(\alpha >0)$, then we may write the factor in Eq.(17)
$f=(\tilde{k}_r+\tilde{h})r-(k_r+h)\tilde{r}=
\alpha^2(1-\alpha^2)m^6_U\{[K_{Ur}(p)+H_U(p)]R_D(p)-[K_{Dr}(p)+H_D(p)]R_U(p)\}$.
In view of Eqs.(14) and (15) as well as the fact that $p_r^2=m_{\phi_S^0}^2$ should
obey the mass inequalities (16), so if $\alpha<1(m_D<m_U)$, then we will have 
$R_U(p)=0$ and obtain $f=\alpha^2(1-\alpha^2)m^6_U[K_{Ur}(p)+H_U(p)]R_D(p)<0$. 
Similarly, if $\alpha>1 \ (m_D>m_U)$ we will have $R_D(p)=0$ and get
$f=-\alpha^2(1-\alpha^2)m^6_U[K_{Dr}(p)+H_D(p)]R_U(p)<0$. As a result, opposite to 
the inference in Ref. [10], we always have $b(p_r)<0$ thus $p_i^0<0$ for positive 
energy $p_r^0$. This means that when  $0\neq m_U\neq m_D\neq 0$, $\phi_S^0$ will 
decay in time instead of the conclusion that its amplitude will grow in time. This 
modification comes from the fact that in present calculation we have carefully 
separated the imaginary part $K_{Qi}(p)$ of the zero-temperature loop integral from 
relevant expressions  e.g. $K_Q(p)$, $S_Q(p)$ and $S_Q^I(p)$ etc. and determined the 
sign of $R_Q(p)$. The same correction is also applicable to the case of $T\to 0$.
When $T=0$, if $m_U\neq m_D$, based on the results that if $p^2<4m_Q^2$ $K_{Qi}(p)=0$ 
and if $p^2\geq 4m_Q^2$ $K_{Qi}>0$ obtained from Eq.(5) by direct calculation 
(instead of $K_{Qi}<0$ by assumption in Ref.[10]) and the similar demonstration to 
above, we may conclude that the amplitude of $\phi_S^0$ will also decay instead grow 
in time. \\
\indent Next we turn to the neutral pseudoscalar bound state $\phi_P^0$. The 
discussion is almost parallel to the one of $\phi_S^0$. In the imaginary-time 
formalism, by keeping the original form of Eq.(1) and using Eq. (6) we may change 
the physical causal propagator for $\phi_P^0$ expressed by Eq.(4.8) in Ref.[10] into  
\begin{equation}
\Gamma_I^{\phi_P^0}(p)
 =-i\frac{\sum_{Q}m_Q^2}{(p^2+i\varepsilon) [k_r+h-ir\sinh(\beta|p^0|/2)]}.
\end{equation}
In the real-time formalism, we only need in Eq. (6.13) in Ref.[10] simply 
to make the replacements $k\to k_r$, $s\to s'$ and $s/r\to s'/r=\cosh(\beta|p^0|/2)$ 
given by Eq.(8) and will obtain correct matrix propagator $\Gamma^{\phi_P^0ba}(p)\; 
(b,a=1,2)$ for $\phi_P^0$. Then diagonalization of $\Gamma^{\phi_P^0ba}(p)\; 
(b,a=1,2)$ by the thermal transformation matrix $M_P$ will lead to the physical causal 
propagator $\Gamma_R^{\phi_P^0}(p)$ for $\phi_P^0$ which is proven to satisfy 
$\Gamma_R^{\phi_P^0}(p)= \Gamma_I^{\phi_P^0}(p)$, i.e. the physical causal 
propagator for $\phi_P^0$ has identical expression in the two formalisms. In 
addition, the derived $M_P$ is equal to $M_S$ given by Eq.(11), thus the thermal 
transformation matrix of the matrix propagator for the neutral pseudoscalar bound 
state $\phi_P^0$ is also the same as the one for an elementary neutral scalar 
particle.\\
\indent Lastly we discuss the propagator for charged scalar bound states $\phi^{\mp}$.  In the imaginary-time formalism, by the analytic continuation of the Matsubara 
frequency 
$\Omega_m$
\begin{equation}
-i\Omega_m +\mu_D-\mu_U\longrightarrow p^0 +i\varepsilon \eta(p^0), \ \ \
                     \varepsilon =0_+,
\end{equation}
we will obtain the physical causal propagator for $\phi^-$ (and $\phi^+$) [10]
\begin{widetext}
\begin{equation}
\Gamma_I^{\phi^-}(p)=-i/ \left\{
(p^2+i\varepsilon)\left[K_{UD}(p)+H_{UD}(p)\right]+
E_{UD}(p)-i(p^2-\bar{M}^2+i\varepsilon) S^I_{UD}(p)\right\},
\end{equation}
where we express alternatively
\begin{equation}
K_{UD}(p)=\frac{1}{p^2+i\varepsilon}\frac{4d_Q(R)}{m_U^2+m_D^2}\int
\frac{id^4l}{(2\pi)^4}\frac{(m_D^2-m_U^2)l\cdot p-m_U^2(p^2+i\varepsilon)}
{(l^2-m_U^2+i\varepsilon)[(l+p)^2-m_D^2+i\varepsilon]}
\end{equation}
which is actually equal to Eq.(5.25) in Ref.[10] and
\begin{equation}
S^I_{UD}(p)=\eta(p^0)4\pi^2 d_Q(R)\int \frac{d^4l}{{(2\pi)}^4}
          \delta(l^2-m_U^2)\delta[(l+p)^2-m_D^2] \left[\sin^2\theta(l^0,\mu_U)\eta(l^0+p^0)+\sin^2\theta(l^0+p^0, \mu_D)\eta(-l^0)\right].
\end{equation}
\end{widetext}
which differs from Eq.(5.28) in Ref.[10] and is the result of rigorously keeping the 
general form of the right-handed side of Eq.(19). By applying the residue theorem 
of complex $l^0$ integral to Eq.(21), we may find out the imaginary part of $K_{UD}(p)$
\begin{equation}
K_{UDi}(p)=\left[1-\frac{p^2}{(p^2)^2+\varepsilon^2}{\bar{M}}^2\right]\Delta_{UD}(p),
\end{equation}
\begin{eqnarray}
\Delta_{UD}(p)&=&\frac{d_Q(R)}{16\pi^2}\int\frac{d^3l}{\omega_{Ul}\omega_{Dl+p}}\left[\delta(p^0+\omega_{Ul}+\omega_{Dl+p})\right.\nonumber\\ &&\left.+\delta(p^0-\omega_{Ul}-\omega_{Dl+p})\right].
\end{eqnarray}
Noting that when $m_U=m_D=m_Q$ we will have $K_{UDi}(p)=\Delta_{UD}(p)$ to be reduced 
to $K_{Qi}(p)$ in Eq.(5). If we explicitly write $K_{UD}(p)= K_{UDr}(p)+i
K_{UDi}(p)$ and use the relation
\begin{equation}
S_{UD}^I(p)-\Delta_{UD}(p)= R_{UD}(p)\eta(p^0)\sinh\frac{\beta(p^0-\mu)}{2},
\end{equation}
where $R_{UD}(p)$ was given by Eq.(6.21) in Ref.[10] and $\mu=\mu_D-\mu_U\equiv
\mu_{\phi^-}$ is the chemical potential of $\phi^-$, then Eq.(20) will be changed 
into
\begin{widetext}
\begin{equation}
\Gamma_I^{\phi^-}(p)=-i/ \left\{
(p^2+i\varepsilon)\left[K_{UDr}(p)+H_{UD}(p)\right]+
E_{UD}(p)-i(p^2-\bar{M}^2+i\varepsilon)R_{UD}(p)\eta(p^0)\sinh
\frac{\beta(p^0-\mu)}{2}\right\}.
\end{equation}
\end{widetext}
In the real-time formalism, it is indicated that in the expression for the matrix 
propagator $\Gamma^{\phi^-ba}(p) \ (b,a=1,2)$ given by Eq.(6.19) in Ref.[10], the 
fact that $K_{UD}(p)$ is complex was ignored. Now if taking this into account and 
noting Eq.(23), we will obtain correct expression for 
$\Gamma^{\phi^-ba}(p) \ (b,a=1,2)$ from Eq.(6.19) in Ref.[10] and successive 
modified results by means of the replacements
\begin{eqnarray}
K_{UD}(p)&\rightarrow &K_{UDr}(p), \nonumber \\ 
S_{UD}(p)&\rightarrow &S^{\prime}_{UD}(p)=S_{UD}(p)-\Delta_{UD}(p)
\end{eqnarray}
and the derived relation 
\[ S^{\prime}_{UD}(p)=R_{UD}(p)\cosh[\beta(p^0-\mu)/2],\]
\begin{equation}
\sqrt{S^{\prime 2}_{UD}(p)-R_{UD}^2(p)}=R_{UD}(p)\eta(p^0)\sinh[\beta(p^0-\mu)/2].
\end{equation}
It is proven that through diagonalization of
$\Gamma^{\phi^-ba}(p) \ (b,a=1,2)$
by the thermal transformation matrix $M_C$ the resulting physical propagator
$\Gamma_R^{\phi^-}(p)$ will have identical form to $\Gamma_I^{\phi^-}(p)$ in 
Eq.(26). This shows equivalence of the two formalisms once again. In addition, $M_C$ 
will have the expression
\begin{widetext}
\begin{equation}
M_C=\left(\matrix{\cosh \theta_C & e^{-\beta \mu/2}\sinh \theta_C\cr
                  e^{\beta\mu/2}\sinh \theta_C &  \cosh \theta_C\cr}\right), \qquad
\sinh\theta_C=\left[\frac{\theta(p^0)}{e^{\beta(p^0-\mu)}-1}+ 
           \frac{\theta(-p^0)}{e^{\beta(-p^0+\mu)}-1}\right]^{1/2}.
\end{equation}
\end{widetext}
Eq.(29) shows that the thermal transformation matrix $M_C$ of the 
matrix propagator for the charged scalar bound state $\phi^-$ with chemical potential 
$\mu$ is identical to the one for an elementary charged scalar particle with chemical 
potential $\mu$ (noting that $M_C$ in Eq. (29) differs from usual one [2] by a 
transpose since our original definition of the matrix $\Gamma_{\phi^-}^{ba}(p) \ 
(b,a=1,2)$ is just so ). \\
\indent In conclusion, by means of keeping general expressions of the analytic 
continuations of the Matsubara frequencies in the imaginary-time formalism and 
separating explicitly the imaginary parts of the zero-temperature loop integrals from 
relevant expressions e.g. $S_Q^I(p)$, $S_Q(p)$, $S_{UD}^I(p)$ and $S_{UD}(p)$ etc., 
we have reanalyzed the results in Ref.[10] and proven identity of the physical causal 
propagators for every scalar bound state in the two formalisms of thermal field theory 
in the one generation NJL model. This dismisses the doubt about possible discrepancy 
between the two formalisms in this problem. Next the derived identity between the 
thermal transformation matrices of the matrix propagators for scalar bound states 
and corresponding elementary scalar particles including the case without and with 
chemical potential indicates similarity of thermodynamic property between these two 
types of particles, even though these bound states could be linear combinations of 
the scalar or pseudoscalar configurations of the Q-fermions with different flavors. 
The reanalysis have not changed the main conclusions of the Nambu-Goldstone mechanism 
at finite temperature reached in Ref.[10] except that the composite Higgs $\phi_S^0$ 
will decay in time instead of its amplitude's growing in time when the two flavors 
of fermions have unequal nonzero masses. 
 
This work was partially supported by the National Natural Science Foundation
of China and by Grant No.LWTZ-1298 of the Chinese Academy of Sciences.


\begin{references}
\bibitem{kn:1}  D. A. Kirzhnits and A. D. Linde, Phys. Lett.  \textbf{42B}, 471
                (1972);
                S. Weinberg, Phys. Rev. D \textbf{7}, 2887 (1973);
                \textbf{9}, 3357 (1974); 
                L. Dolan and R. Jackiw, \textit{ibid.} \textbf{9}, 3320 (1974);
                A. D. Linde, Rep. Prog. Phys. \textbf{42}, 389 (1979);
                L. Girardello, M. T. Grisaru, and P. Salomonson, Nucl. Phys.
                \textbf{B178}, 331 (1981);
                B. deWitt, in:\textit {Fundamental Interactions}, Carg\`{e}se 1981,
                edited by. M. Levy \textit {et al.} (Plenum, New York, 1982);
                R. H. Brandenberger, Rev. Mod. Phys. \textbf {57}, 1 (1985);
H. Umezawa, H. Matsumoto, and M. Tachiki, \textit{Thermo-field
                dynamics and condensed matter states} (North-Holland, Amsterdam,
                1982);
                K. C. Chou, Z. B. Su, B. L. Hao and L. Yu, Phys. Rep. 
                \textbf {118}(1985)1.
\bibitem{kn:2}  N. P. Landsman and Ch. G. van Weert, Phys. Rep. \textbf{145},
                141 (1987). 
\bibitem{kn:3}  J. I. Kapusta, \textit{Finite-temperature field theory}, (Cambridge University Press, Cambridge, England, 1989);
                B. Rosenstein, B. J. Warr, and S. H. Park, Phys. Rep. \textbf{205},
                59 (1991); 
                M.L. Bellac,  \textit{Thermal Field Theory}( Cambridge University
                Press, Cambridge, England, 1996). 
\bibitem{kn:4}  Y. Fujimoto, R. Grigjanis and H. Nishino,  Phys. Lett. 
                \textbf{141B} (1984) 83;
                Y. Fujimoto and R. Grigjanis, Z. Phys. C \textbf{28} (1985) 395;
                      Prog. Theor. Phys. \textbf{74} (1985) 1105;
                R. Kobes, Phys. Rev. D \textbf{42} (1990) 562; \textbf{43} (1991) 1269.
                E. Braaten and R. Pisarski, Phys. Rev. Lett. \textbf{64} (1990) 1338;                      Nucl. Phys. B \textbf{337} (1990) 569;
                J. Frenkel and J.C. Taylor, Nucl. Phys. B \textbf{334} (1990) 199. 
\bibitem{kn:5}  R. L. Kobes and G.W. Semenoff, Nucl. Phys. B \textbf{260}, 714 (1985);                 \textit{ibib.} B \textbf{272}, 329 (1986);
                Y. Fujimoto, M. Morikana and M. Sasaki, Phys. Rev. D \textbf{33}, 
                590(1986).
\bibitem{kn:6}  T. S. Evans, Phys. Lett. B \textbf{249}, 286 (1990); \textbf{252}, 108 (1990).  
\bibitem{kn:7}  R. Kobes, Phys. Rev. Lett. \textbf{67}, 1384 (1991); 
                M. A. van Eijck, R. Kobes and Ch. G. van Weert, Phys. Rev. 
                D \textbf{50}, 4097 (1994). 
\bibitem{kn:8}  Y. Nambu, Phys. Rev. Lett. \textbf{4}, 380 (1960);
                J. Goldstone, Nuovo Cimento \textbf{19}, 154 (1961);
                J. Goldstone, A. Salam, and S. Weinberg, Phy. Rev. \textbf{127},
                965 (1962);
                S. Bludman and A. Klein, \textit{ibid.}  \textbf{131}, 2363 (1962).
\bibitem{kn:9}  Y. Nambu and G. Jona-Lasinio, Phys. Rev. \textbf{122}, 345 (1961);
                \textbf{124}, 246 (1961).                                      
\bibitem{kn:10}  B. R. Zhou, Phys. Rev. D \textbf{62}, 105004 (2000). 
\bibitem{kn:11}  B. R. Zhou, Phys. Rev. D \textbf{59}, 065007 (1999).
\bibitem{kn:12}  B. R. Zhou, Phys. Rev. D \textbf{57}, 3171 (1998); 
                 Commun. Theor. Phys. \textbf{32}, 425 (1999).  
\end{references}
\end{document}